\documentclass[tighten,iop,apj]{emulateapj}
\pdfoutput=1
\usepackage{amssymb}
\usepackage{amsfonts}
\usepackage{amsmath}
\usepackage[varg]{txfonts}
\usepackage{natbib}
\usepackage{color}
\usepackage{graphicx}
\usepackage{hyperref}
\usepackage{aas_macros}

\newcommand{\pd}{\partial}
\newcommand{\ud}{\ensuremath{\mathrm{d}}}

\begin{document}

\title{An FFT-based Solution Method for the Poisson Equation on 3D Spherical Polar Grids}

\author{Bernhard~M\"uller and Conrad~Chan}
\affiliation{School of Physics and Astronomy, Monash University, Clayton, VIC~3800, Australia}

\begin{abstract}
The solution of the Poisson equation is a ubiquitous
problem in computational astrophysics. Most notably,
the treatment of self-gravitating flows 
involves the Poisson equation for the
gravitational field. In hydrodynamics codes using 
spherical polar grids, one often resorts to a 
truncated spherical harmonics expansion for an 
approximate solution. Here we present a 
non-iterative
method that is similar in spirit, but uses
the full set of eigenfunctions of the discretized
Laplacian to obtain an exact solution
of the discretized Poisson equation. 
This allows the solver to handle
density distributions for which the truncated
multipole expansion fails, such as off-center
point masses.
In three dimensions, the operation
count of the new method is competitive with a naive 
implementation of the truncated spherical
harmonics expansion with $N_\ell \approx 15$
multipoles.
We also discuss the parallel
implementation of the algorithm. The serial code and a template for the parallel solver are made publicly available.
\end{abstract}


\section{Introduction}
\label{sec:intro}
The numerical solution of the Poisson equation is one of the standard
problems in astrophysical fluid dynamics. The Poisson equation is
probably encountered most frequently as the equation governing the
gravitational field in the Newtonian approximation, but its
applications also include constrained formulations of general
relativity \citep[e.g.][]{cordero_09}, projection methods for
magnetohydrodynamics \citep{brackbill_80,leveque}, anelastic/low-Mach
number flow \citep{batchelor_53,ogura_62,jacobson_99}, and radiation transport problems
\citep{liebendoerfer_09}.

Various methods for the exact or approximate solution of the Poisson
equation are commonly used in astrophysical codes.  The applicability
and usefulness of these methods is typically dictated by the
geometry of the physical problem at hand and the discretization
technique used for the equations of hydrodynamics.  In stellar
hydrodynamics approximate spherical symmetry obtains, so that
spherical polar grids (including overset grids,
\citealp{wongwathanarat_10a}) are often the method of choice. For such
grids, fast algorithms such as the direct use of the three-dimensional
Fast Fourier Transform (FFT) \citep{hockney_65,eastwood_79}, 
multi-grid algorithms \citep{brandt_77}, or tree algorithms
\citep{barnes_86} are either not directly applicable, more difficult to
implement, or do not offer a good trade-off between computational
efficiency and accuracy. One of the most frequently used methods for
``star-in-a-box'' simulations has long been based on a spherical
harmonics expansion of the Green's function as described by
\citet{mueller_95}. Since the gravitational field typically deviates
only modestly from spherical symmetry for such problems, the spherical
harmonics expansion can be truncated at a low multipole number
$N_\mathrm{\ell}=10\ldots 20$ for better  computational
efficiency. The overall operation count of the algorithm is only
$\mathcal{O}(N_r N_\theta N_\ell)$ for a spherical polar grid with
$N_r\times N_\theta$ zones in the $r$- and $\theta$-direction in the case
of axisymmetry (2D), and $\mathcal{O}(N_r N_\theta N_\varphi N_\ell^2)$ in
three dimensions (3D) with $N_\varphi$ zones in the
$\varphi$-direction. The high efficiency of the algorithm has made it
the method of choice for several supernova codes employing spherical
polar grids such as the the \textsc{Chimera} code \citep{bruenn_13},
the \textsc{Aenus} code \citep{obergaulinger_06a}, the \textsc{Fornax}
code \citep{burrows_18}, and various offshoots of the
\textsc{Prometheus} code \citep{marek_09,wongwathanarat_10a}.  The
method has also been adapted \citep{couch_13b} for simulations of
stellar hydrodynamics problems using the \textsc{Flash} code
\citep{fryxell_00}.

Despite its efficiency, this algorithm still has some drawbacks.
Above all, it only offers an approximate solution to the
Poisson equation. Although the error is usually acceptable when the
algorithm is used to obtain the gravitational field, this precludes
its use, e.g., for divergence cleaning in the MHD projection method, which
requires an exact solution of the discretized Poisson equation. 
An exact solution is also desirable if one seeks to implement
gravitational forces in a momentum-conserving form \citep{shu,livne_04}
and can be exploited to achieve total energy conservation to machine precision
\citep{mueller_10}.  The truncation of the spherical harmonics
expansion is especially problematic when the location of the central
density peak of the source does not coincide with the origin of the
coordinate system. Although this can be fixed by a judicious choice of
the center of the multipole expansion \citep{couch_13b}, such a fix
destroys much of the simplicity of the algorithm in spherical polar
coordinates. Finally, there are subtle problems with the convergence
of the multipole expansion.  \citet{couch_13b} noted that a naive
implementation of the algorithm can include a spurious
self-interaction term that manifestly leads to divergence for large
$N_\ell$. This can again be fixed -- either by the original method of
\citet{mueller_95} or that of \citet{couch_13b} -- but more subtle
problems still lurk when one projects the source
density onto spherical harmonics: Analytically, one
has the orthogonality relation
\begin{equation}
\int Y^*_{\ell m} Y_{\ell' m'} \,\ud \Omega
=\delta_{\ell \ell'}\delta_{m m'},
\end{equation}
which implies that the gravitational field $\Phi$ only contains
exactly the same multipole components as the source. This is generally
not the case for the discretized integrals. Though the orthogonality
relation is easily maintained if either $\ell=0$ or $\ell'=0$, and for
multipoles of opposite parity, multipoles with $\ell\ge 1$ in the
density field will generally give rise to spurious multipoles of
\emph{arbitrarily high} $\ell$.  This spurious overlap between spherical
harmonics of different $\ell$ and $m$ only vanishes in the limit of
infinite spatial resolution. This problem is illustrated further in the Appendix.

In this article, we point out that all of these problems can be
avoided by solving the discretized Poisson equation \emph{exactly}
using the discrete analogue of the spherical harmonics expansion in
conjunction with the FFT in the $\varphi$-direction.
The approach of combining the FFT with
Legendre and Chebyshev transforms to exactly invert the
Poisson equation is well established in
the field of pseudospectral methods
\citep[e.g.][]{fornberg_95,chen_00,lai_02,weatherford_05}.
Such pseudospectral approaches
can only be combined
with finite-volume methods at a cost, however. Since
the collocation points of the pseudospectral
grid generally differ from the finite-volume
grid, mapping is required, which can impact
performance and impede parallelization. Moreover,
the superior accuracy of pseudospectral methods
for the elliptical part of the problem
is typically of little advantage when the
error is mainly determined by the hyperbolic
finite-volume solver; in those cases 
consistency between the elliptic and hyperbolic
solver and the enforcement of physical conservation
laws is often a higher priority than the nominal
accuracy of the elliptic solver. For these reasons,
we here construct an exact algorithm in the
flavor of pseudospectral codes that works 
on the finite-volume grid itself.

The operation
count of our algorithm remains competitive with the method of
\citet{mueller_95} in 3D; for an angular resolution of $N_\theta\times
N_\varphi=128\times 256$, the break-even point of the serial algorithm
is at $\ell_\mathrm{max}\approx 15$.  Although
the mathematics behind the algorithm is simple and merely based on
standard methods from the theory of partial differential equations and
linear algebra, it is not currently used in astrophysical fluid
dynamics codes and no off-the-shelf implementation is available.
Along with the paper, we therefore provide the code of the serial
implementation, which uses the \textsc{Lapack} \citep{anderson_99} and
\textsc{FFTW} \citep{frigo_05} libraries, and a template for the parallel
version with domain decomposition in $\theta$ and $\varphi$.

Our paper is structured as follows: As a preparation
for the solution of the discretized Poisson equation,
we recapitulate how the multipole expansion of
\citet{mueller_95} can
be obtained directly by separation of variables. We
then formulate the discrete analogue of
the multipole expansion in Section~\ref{sec:discrete}
and also discuss its parallelization.
In Section~\ref{sec:performance} we discuss the efficiency
of the serial and parallel version of the algorithm,
then proceed to code verification in Section~\ref{sec:verification}, and
 end with a brief summary in Section~\ref{sec:summary} 

\section{Multipole Expansion by Separation
of Variables}
\label{sec:multipole}
The algorithm of \citet{mueller_95} for the solution 
of the Poisson equation for
the gravitational potential $\Phi$ and the source
density $\rho$,
\begin{equation}
\Delta \Phi =4 \pi G \rho,
\end{equation}
is usually derived by writing the solution in terms
of the Green's function $\mathcal{G}$, as
\begin{equation}
\label{eq:green}
 \Phi(\mathbf{r}) =-G\int \mathcal{G}(\mathbf{r}-\mathbf{r'}) \rho(\mathbf{r'})
\,\ud^3 \mathbf{r}'.
\end{equation}
The Green's function is given by $\mathcal{G}(\mathbf{r}-\mathbf{r'})
= |\mathbf{r}-\mathbf{r'}|^{-1}$, and can be expanded in terms of
spherical harmonics $Y_{\ell m}$ as
\begin{eqnarray}
\label{eq:green2}
\mathcal{G}(\mathbf{r},\mathbf{r'})&=&\frac{1}{|\mathbf{r}-\mathbf{r'}|}
\\
\nonumber
&=&\sum_{\ell=0}^\infty \sum_{m=-\ell}^\ell
\frac{4\pi}{2 \ell+1}Y_{\ell m}(\theta,\varphi)Y^*_{\ell m}(\theta',\varphi')
\frac{\min(r,r')^\ell}{\max(r,r')^{\ell+1}}.
\end{eqnarray}
After inserting this expansion
into Equation~(\ref{eq:green})
and projecting out the individual multipole components, one can obtain individual multipoles
$f_{\ell,m}$ of the solution by 
integration along the radial direction,
\begin{eqnarray}
\label{eq:sol_r}
f_{\ell,m}(r)&=&
-\frac{4\pi}{2 \ell+1}
\left[\frac{1}{r^{\ell+1}}\int_{0}^r \bar{\rho}_{\ell,m}(r') r'^{\ell+2} \, \ud r'
\right.
\\
\nonumber
&&
\left.
+
r^\ell \int_{r}^\infty \bar{\rho}_{\ell,m}(r') r'^{\ell+2} \, \ud r'
\right],
\end{eqnarray}
and then reconstruct the full solution as
\begin{equation}
\Phi=G \sum_{\ell=0}^\infty \sum_{m=-\ell}^\ell f_{\ell,m}(r)
Y_{\ell m} (\theta,\varphi).
\end{equation}
Here $\bar{\rho}_{\ell,m}(r)$ are the multipoles of the source
density.

In fact, there is no need to ever invoke the explicit form
$\mathcal{G}(\mathbf{r}-\mathbf{r'}) = |\mathbf{r}-\mathbf{r'}|^{-1}$
of the Green's function and the specific expansion in
Equation~(\ref{eq:green2}) to derive this solution:
Instead, one can directly obtain decoupled
ordinary differential equations for $f_{\ell,m}$
by noting that the spherical harmonics
are eigenfunctions of the
angular part $\Delta_\Omega$ of the Laplacian
in spherical polar coordinates,
\begin{equation}
\Delta_\Omega
=
\frac{\pd}{\pd \theta}\left(\sin\theta \frac{\pd }{\pd \theta}\right)
+\frac{1}{\sin^2 \theta}\frac{\pd^2 }{\pd \varphi^2}.
\end{equation}
Using $\Delta_\Omega Y_{\ell m} (\theta,\varphi)=-\ell(\ell+1) 
Y_{\ell m} (\theta,\varphi)$ after inserting the expansion
$\Phi= \sum_{\ell,m} f_{\ell,m}(r) Y_{\ell m} (\theta,\varphi)$
into the Poisson equation then immediately yields
decoupled equations for
the $f_{\ell,m}$,
\begin{equation}
\label{eq:ode}
\frac{1}{r^2}\frac{\pd}{\pd r}\left(r^2\frac{\pd f_{\ell,m}}{\pd r}\right)
-\frac{\ell (\ell+1)}{r^2} f_{\ell,m}
=4\pi G \int Y_{\ell,m}^\star\rho(r,\theta,\varphi) \ud \Omega,
\end{equation}
which can be solved according to Equation~(\ref{eq:sol_r}).
The spherical harmonics themselves
are obtained in an analogous manner by first solving
the eigenvalue problem for the azimuthal
part 
$\Delta_\varphi=\pd^2 /\pd \varphi^2$
of the Laplacian
and then solving another set of eigenvalue problems
for $\Delta_\Omega$.

\section{Solution of the Discrete Poisson Equation}
\label{sec:discrete}
For the discretized Poisson equation, one can apply a completely
analogous procedure to first obtain the eigenvectors of the $\varphi$-derivative terms in the discrete
Laplacian, then the eigenfunctions for $\Delta_\Omega$, and finally
decoupled equations for the radial dependence of the individual
multipole components.

\subsection{Discretisation}
We discretize the Poisson equation as
\begin{equation}
(\Delta \Phi)_{i,j,k}=
(\Delta_r \Phi)_{i,j,k}+
(\Delta_\theta \Phi)_{i,j,k}+
(\Delta_\varphi \Phi)_{i,j,k}
=s_{i,j,k}
\label{eq:fin_vol0}
\end{equation}
where the source is $s_{i,j,k}=4\pi G \rho_{i,j,k}$
and
$i$, $j$, and $k$ are the grid indices in the $r$-, $\theta$-,
and $\varphi$-direction. Values offset by $1/2$ will be used
for
quantities at cell interfaces. 
The discretized
operators $\Delta_r$, $\Delta_\theta$, and $\Delta_\varphi$
for the $r$-, $\theta$, and $\varphi$-derivatives are
\begin{eqnarray}
\nonumber
(\Delta_r \Phi)_{i,j,k} &=& 
\frac{1}{\delta V_{i,j,k}}
\left(
\frac{\Phi_{i+1,j,k}-\Phi_{i,j,k}}{r_{i+1}-r_i} \delta A_{i+1/2,j,k} 
\right.
\\
&&
\left.
-
\frac{\Phi_{i,j,k}-\Phi_{i-1,j,k}}{r_{i}-r_{i-1}} \delta A_{i-1/2,j,k} 
\right),
\\
\nonumber
(\Delta_\theta \Phi)_{i,j,k} &=& 
\frac{1}{\delta V_{i,j,k}}
\left(
\frac{\Phi_{i,j+1,k}-\Phi_{i,j,k}}{r_i \left(\theta_{j+1}-\theta_{j}\right)} \delta A_{i,j+1/2,k} 
\right.
\\
&&
\left.
-
\frac{\Phi_{i,j,k}-\Phi_{i,j-1,k}}{r_i \left(\theta_{j}-\theta_{j-1}\right)} \delta A_{i,j-1/2,k} 
\right),
\\
\nonumber
(\Delta_\varphi \Phi)_{i,j,k} &=& 
\frac{1}{\delta V_{i,j,k}}
\left(
\frac{\Phi_{i,j,k+1}-\Phi_{i,j,k}}{r_{i} \sin \theta_j \left(\varphi_{k+1}-\varphi_{k}\right)} \delta A_{i,j,k+1/2} 
\right.
\\
&&
\left.
-
\frac{\Phi_{i,j,k}-\Phi_{i,j,k-1}}{r_{i} \sin \theta_j \left(\varphi_{k+1}-\varphi_{k}\right)} \delta A_{i,j,k-1/2} 
\right).
\label{eq:fin_vol3}
\end{eqnarray}
Here, $\delta V$ and $\delta A$ denote cell volumes and interface
areas, respectively.  We note that this is a second-order accurate
(for uniform grids in $r$, $\theta$, and $\varphi$) finite-volume
discretisation of the integral form $\oint \nabla \Phi \cdot
\mathbf{dA} = 4 \pi G \int \rho \,\ud V$ of the Poisson equation,
which allows us to write the energy source term in the Newtonian
equations of hydrodynamics such that total energy is conserved to
machine precision \citep{mueller_10}.

In order to utilize the FFT in the solution algorithm,
we require a uniform grid in $\varphi$ with spacing $\delta \varphi$.
For the sake of simplicity, we also use uniform grid spacing
in the $\theta$-direction, although this is not required
for a solution by separation of variables. In this case,
one obtains the following interface surfaces and cell volumes
by analytic integration,
\begin{eqnarray}
\delta A_{i+1/2,j,k}
&=&
r_{i+1/2}^2
(\cos \theta_{i,j-1/2,k}-\cos \theta_{i,j+1/2,k})
\,\delta \varphi,
\\
\delta A_{i,j+1/2,k}
&=&
\frac{r_{i+1/2}^2-r_{i+1/2}^2}{2}
\sin \theta_{j+1/2}
\,\delta \varphi,
\\
\delta A_{i,j,k+1/2}
&=&
\frac{r_{i+1/2}^2-r_{i+1/2}^2}{2}
\,\delta \theta,
\\
\delta V_{i,j,k}
&=&
\frac{r_{i+1/2}^3-r_{i-1/2}^3}{3}
\\
&&
\nonumber
\times(\cos \theta_{i,j-1/2,k}-\cos \theta_{i,j+1/2,k})
\,\delta \varphi.
\end{eqnarray}

Before proceeding further, it is convenient to factor out terms that
depend on $r$ in $\Delta_\theta$ and on $r$ and $\theta$ in
$\Delta_\varphi$. We therefore define new operators
$\hat{\Delta}_\theta$ and
$\hat{\Delta}_\varphi$ such that
\begin{eqnarray}
(\Delta_\theta \Phi)_{i,j,k} &=& 
\mathcal{R}_i
\left(
\frac{\Phi_{i,j+1,k}-\Phi_{i,j,k}}{\delta \theta} \sin \theta_{j+1/2}
\right.
\\
\nonumber
&&
\left.
-
\frac{\Phi_{i,j,k}-\Phi_{i,j-1,k}}{\delta \theta} \sin \theta_{j-1/2}
\right),
\\
\nonumber
&=&\mathcal{R}_i (\hat{\Delta}_\theta \Phi)_{i,j,k}
\\
(\Delta_\varphi \Phi)_{i,j,k} &=& 
\mathcal{R}_{i}
\mathcal{S}_j
\left(
\frac{\Phi_{i,j,k+1}-\Phi_{i,j,k}}{\delta \varphi}
-
\frac{\Phi_{i,j,k}-\Phi_{i,j,k-1}}{\delta \varphi}
\right)
\nonumber
\\
&=&
\mathcal{R}_{i}
\mathcal{S}_j
(\hat{\Delta}_\varphi \Phi)_{i,j,k},
\end{eqnarray}
where
\begin{eqnarray}
\mathcal{R}_i
&=&
\frac{3(r_{i+1/2}^2-r_{i-1/2}^2)}
{2r_i (r_{i+1/2}^3-r_{i-1/2}^3)},
\\
\mathcal{S}_j
&=&
\frac{\delta \theta}
{\sin \theta_j (\cos \theta_{i,j-1/2,k}-\cos \theta_{i,j+1/2,k})}.
\end{eqnarray}

\subsection{Description of the Serial Algorithm}

To solve the discretized Poisson equation, we first expand
the solution in terms of the eigenvectors
of $\hat{\Delta}_\varphi$
The eigenvectors $h_m(k)$ and eigenvalues $\lambda_{\varphi,m}$
are given by
\begin{equation}
h_m (k) = e^{2 \pi i m k / N_\varphi} ,
\quad
\lambda_{\varphi,m} 
=
 \left(\frac{\sin m\, \delta \varphi/2}{\delta \varphi/2}\right)^2 
,
\end{equation}
where $m$ can take on values between $0$ and $N_\varphi-1$.  Expressing both
$\Phi$ and the source $s$ in terms of the eigenfunctions and Fourier
components $g_{i,j,m}$ and $\tilde{s}_{i,j,m}$,
\begin{equation}
\Phi_{i,j,k}=
\sum_{m}^{N_\varphi-1} g_{i,j,m} h_m(k),
\end{equation}
\begin{equation}
s_{i,j,k}=
\sum_{m=0}^{N_\varphi-1} \tilde{s}_{i,j,m} h_m(k),
\end{equation}
yields 
\begin{eqnarray}
\nonumber
&&\sum_{m=0}^{N_\varphi-1}
\left[(\Delta_r g)_{i,j,m}+
\mathcal{R}_i(\hat{\Delta}_\theta g)_{i,j,m}+
  \mathcal{R}_i \mathcal{S}_j
\lambda_{\varphi,m} g_{i,j,m}\right] h_m(k)
\\
&&=\sum_{m=0}^{N_\varphi-1}
\tilde{s}_{i,j,m} h_m(k).
\end{eqnarray}
Projecting on the orthogonal eigenvectors
yields a partially decoupled system
of equations for $g_{i,j,m}$,
\begin{equation}
(\Delta_r g)_{i,j,m}+
\mathcal{R}_i
\left[(\hat{\Delta}_\theta g)_{i,j,m}+
 \mathcal{S}_j\lambda_{\varphi,m} g_{i,j,m}
\right]
=
\tilde{s}_{i,j,m}.
\end{equation}
Here $\tilde{s}_{i,j,m}$ can be obtained efficiently
from $s_{i,j,k}$  using the FFT.

To fully decouple the system, we expand $g$ and the $\tilde{s}$
further in terms of the orthonormal eigenvectors of the operator
$\hat{\Delta}_\theta+\mathcal{S}_j \lambda_{\varphi,m}$, i.e.\ in
terms of the $N_\theta$ vectors $H_{\ell,m}(j)$ that fulfill
\begin{equation}
(\hat{\Delta}_\theta H_{\ell,m})(j)
+ \mathcal{S}_j\lambda_{\varphi,m} H_{\ell,m}(j)
=
\lambda_{\Omega,\ell,m} H_{\ell,m}(j).
\end{equation}
Although the computation of the complete set of eigenvectors
for each $m$ can be expensive, it only needs to
be carried out once in an Eulerian code when the solver
is set up.

Expanding
\begin{equation}
\label{eq:leg_forward}
g_{i,j,m}
=
\sum_{\ell=0}^{N_\theta-1}
f_{i,\ell,m} H_{\ell,m}(j),
\end{equation}
and projecting onto $H_{\ell,m}$ gives
\begin{equation}
\label{eq:tridiag}
\Delta_r f_{i,\ell,m}+
 \mathcal{R}_i\lambda_{\Omega,\ell,m} 
=
\hat{s}_{i,\ell,m}.
\end{equation} 
Transforming $\tilde{s}$ to $\hat{s}$ now involves a
matrix-vector multiplication with the inverse of the matrix
$H_{\ell,m}(j)$. 

Equation~(\ref{eq:tridiag}) amounts to a set of decoupled boundary
value problems. For each $\ell$ and $m$, a tridiagonal linear system
needs to be solved. The outer boundary condition is best implemented
at this stage to ensure compatibility with the analytic
solution in an infinite domain. Analytically,
$f_{\ell,m}(r)$ is found to decrease as
\begin{equation}
f_{\ell,m}(r) \propto
r^{-\frac{1+\sqrt{1-4\lambda_\Omega}}{2}}
\end{equation}
at large distances from the sources.
This suggests that we replace the finite-difference
approximation
for the derivative of $f$ at the outer boundary
with the extrapolated value of $\pd f_{\ell,m}/\pd r$
using the value of $f_{\ell,m}$ in the outermost zone on
the grid,
\begin{eqnarray}
\frac{f_{N_r+1,\ell,m}-f_{N_r+1,\ell,m}}{r_{N_r+1}-r_{N_r}}
&\rightarrow&
-\frac{1+\sqrt{1-4\lambda_\Omega}}{2}
\frac{f_{N_r+1,\ell,m}}{r_{N_r+1/2}}
\nonumber
\\
&&\times
\left(\frac{r_{N_r+1/2}}{r_{N_r}}\right)^{-\frac{1+\sqrt{1-4\lambda_\Omega}}{2}}.
\end{eqnarray}

Once $f_{i,\ell,m}$ has been determined, one obtains
$g_{i,j,m}$ by matrix-vector multiplication using
the eigenvector matrix $H_{\ell,m}(j)$, and
then $\Phi_{i,j,k}$ from $g_{i,j,m}$ by means
of another FFT.

\subsection{Parallel Implementation}
\label{sec:parallel}
Both the FFT and the matrix-vector multiplication can be parallelised
used standard domain-decomposition techniques. In principle, libraries
such as \textsc{FFTW3}\footnote{\url{www.fftw.org}} \citep{frigo_05}
for the FFT and
\textsc{Scalapack}\footnote{\url{www.netlib.org/scalapack}}
\citep{choi_95} for the matrix-vector multiplication can be
employed. For better conformance with existing data structures, we
have, however, written our own MPI-parallel version of these
operations to include the exact solver in the relativistic radiation
hydrodynamcis code \textsc{CoCoNuT-FMT} \citep{mueller_15a}, where the
solver is used for obtaining the space-time metric in the extended
conformal flatness approximation of \citet{cordero_09}. 
We use domain decomposition in the $\theta$- and $\varphi$-direction
with a Cartesian topology, and restrict ourselves to
cases where the number of domains $n_\theta\times n_\varphi$ in both directions
is a power of 2.
Standard
\textsc{Lapack}\footnote{\url{www.netlib.org/lapack}}
\citep{anderson_99} and \textsc{Blas}\footnote{\url{www.netlib.org/blas}}
\citep{blackford_02} routines are used for the
determination of eigenvectors, the node-local
part of matrix-vector multiplications, and tridiagonal solves.

The parallelization of the FFT is trivial, and merely requires
point-to-point communication at the appropriate
points in the butterfly diagram. Parallel matrix-vector multiplication is 
implemented as follows: Consider the transformation
from $f_{i,\ell,m}$
to $g_{i,j,m}$ (Equation~\ref{eq:leg_forward}),
\begin{equation}
\label{eq:leg_forward2}
g_{i,j,m}=\sum_{\ell=0}^{N_\theta-1} H_{\ell,m}(j) f_{i,\ell,m}.
\end{equation}
If we suppress
the indices $i$ and $m$, we can write this in the form
\begin{equation}
\mathbf{y}_{J,K}=\sum_{K=1}^{n_\theta} \mathbf{M}_{J,K} \mathbf{x}_{K},
\end{equation}
where the indices $J$ and $K$ run over $n_\theta$ domains in the
$\theta$-direction, and the elements of the matrix $\mathbf{M}$ and
the vectors $\mathbf{x}$ and $\mathbf{y}$ are blocks of size
$(N_\theta/n_\theta)\times (N_\theta/n_\theta)$ and
$N_\theta/n_\theta$.

On any MPI task $J$, all the matrix elements $\mathbf{M}_{J,K}$ are
available, but only one component of the vector $\mathbf{x}$ is. We
can, however, compute $\mathbf{M}_{K,J} \mathbf{x}_J$ for all $K$ on
task $J$. Thus, all the terms appearing in the matrix-vector product
are available right away, but need to be reshuffled between the
different tasks to assemble the dot products between the rows of the
matrix $\mathbf{M}$ and the vector $\mathbf{x}$.

To describe how the terms $\mathbf{M}_{K,J} \mathbf{x}_J$ are exchanged
between different MPI tasks, we introduce the shorthand notation
$\mathcal{P}_{J,\sigma}$ to denote the partial sum
$\sum_{K\in \sigma} \mathbf{M}_{J,K} \mathbf{x}_{K}$. Initially, task $J$
has $\mathcal{P}_{J,\sigma}$ available only for $\sigma=\{J\}$, but
for any $J$. In the end, we require
 $\mathcal{P}_{J,\sigma}$ for $\sigma=\{1,\ldots, n_\theta\}$,
but only for one (local) value of the index $J$. This is accomplished iteratively.
In step $s$ of the iteration, we 
\begin{enumerate}
\item exchange data with task $J+2^{s-1}$ if the $s$-th digit from the right
in the binary representation of $J$ is even and with task $J-2^{s-1}$ if the
$s$-th digit is odd,
\item compute new partial sums $P_{K,\sigma'_j}=P_{K,\sigma_J}+P_{K,
  \sigma_{J\pm 2^{s-1}}}$, which implies that the new $\sigma'$ for
  task $J$ is $\sigma'_J=\sigma_J \cup \sigma_{J\pm 2^{s-1}}$,
\item
    compute and retain those sums only for those $K$ that agree with
    $J$ in the smallest $s$ binary digits.
\end{enumerate}
After the first step, task $J$ holds partial sums
$\mathcal{P}_{K,\sigma}$ only for those $K$ that \emph{agree in the
  last binary digit} with $J$. $\sigma$, on the other hand, is now
larger, and contains all numbers that agree with $J$ \emph{up to and
  excluding the last binary digit}. Subsequent steps further decimate
the partial sums and build up $\sigma$. After step $s$, task $J$ holds
$\mathcal{P}_{K,\sigma}$ for all $K$ that agree with $K$ in the
smallest $s$ binary digits, and $\sigma$ contains all number that
agree with $J$ up to and excluding the smallest $s$ binary
digits. Thus, after $\log_2 n_\theta$ communication steps, task $J$
only holds $\mathcal{P}_{J,\sigma}=\sum_{K=1}^{N_\theta}
\mathbf{M}_{JK} \mathbf{x}_K=\mathbf{y}_J$, i.e.\ each task holds one
local component of the result vector.

\section{Efficiency of the Algorithm}
\label{sec:performance}
\subsection{Serial Version}
It is instructive to compare the operation count for
the exact solver with the algorithm of
\citet{mueller_95}. 

In 2D, the truncated Green's function expansion using
$N_\ell$ multipoles as implemented by \citet{mueller_95} requires
roughly $2 N_r N_\theta N_\ell$ operations, mostly for computing
the multipoles of the source and reconstructing the potential from its
multipoles. In 3D, one has $2\ell+1$ spherical harmonics with
different magnetic quantum number $m$
for each $\ell$, and hence $N_\ell^2$ basis
functions in total. Thus the operation count increases to $2 N_r N_\theta
N_\varphi N_\ell^2$.

The exact solver requires roughly $2 N_r N_\theta N_\varphi
\log_2 N_\varphi$ operations for FFTs, and $2 N_r N_\theta^2
N_\varphi$ for matrix-vector multiplications. Hence the total
operation count is about $2 N_r N_\theta^2$ in 2D and $2 N_r
N_\theta N_\varphi (N_\theta+\log_2 N_\varphi)$.

While the exact solver is invariably more expensive in 2D, it actually
compares favourably to the method of \citet{mueller_95} if
$N_\ell\gtrsim \sqrt{N_\theta}$.  Since one typically needs to account
for at least $N_\ell \gtrsim 10$ multipoles, the exact solver is
competitive for typical grid resolutions of $N_\theta=128\ldots 256$
in core-collapse supernova simulations and outperforms the
straightforward implementation of the truncated spherical harmonics
expansion for $N_\ell \gtrsim 12 \ldots 16$. The truncated moment
expansion could, however, be brought down to $2 N_r N_\theta N_\varphi
\log_2 N_\varphi$ operations if the projection of the density onto
spherical harmonics is broken apart into a projection on Fourier modes
and on associated Legendre polynomials in separate steps,
and if the FFT is used for the
transforming between $\varphi$-space and $m$-space.

\subsection{Parallel Version}
While the computational efficiency of the exact solver is roughly on
par with the truncated spherical harmonics expansion in serial mode,
achieving high parallel performance is more challenging. The reason
for this is the large amount of data that needs to be exchanged
between MPI tasks, mostly for parallel matrix-vector multiplication
(although the cost of communication in the FFT is not negligible either).
The total number of (complex) array elements that are sent in the
first and most expensive step of the multiplication algorithm
described in Section~\ref{sec:parallel} by all tasks combined is $N_r
N_\theta N_\varphi (N_\theta/n_\theta)$.  The subsequent steps of the
algorithm add another factor of 2, and two multiplications per solve
are needed, so that the total amount of data sent scales as $4 N_r
N_\theta N_\varphi (N_\theta/n_\theta)$.

If the same domain decomposition is used for the truncated spherical
harmonics expansion, the amount of data sent during during the
required global reduction operation is only $\mathcal{O}(N_r n_\theta
n_\varphi N_\ell^2)$.

\begin{table}
\centering
 \begin{tabular}{ccc}
 number & \multicolumn{2}{c}{ wall clock tims [s]}\\
 of cores & original & parity-split \\
 \hline\hline
 32 & 0.32 & 0.44\\
 64 & 0.22 & 0.17\\
 128 & 0.18 & 0.12\\
 256 & 0.13 & 0.083\\
 512 & 0.12 & 0.073\\
 1024 & 0.08 & 0.036\\
 \end{tabular}
\caption{Wall-clock time
for a single call to the exact Poisson solver
for different numbers of cores for
a grid of $N_r\times N_\theta\times N_\varphi=
550\times 128 \times 256$ zones
for the unmodified solver (second column)
and for the version that splits the solution
into  components of opposite parity (third column).
\label{tab:scaling}}
\end{table}

For representative values of $N_\theta=128$, $N_\varphi=256$ and
$N_\ell =15$ and several hundred MPI tasks, the volume of the
transmitted data is larger by about one order of magnitude than for
the truncated spherical harmonics expansion. Consequently, the scaling
of the exact algorithm is not optimal as can be seen from the result
of strong scaling tests conducted on \textsc{Magnus} at the Pawsey
Supercomputing Centre (Table~\ref{tab:scaling}).  In contrast to this,
\citet{almanstoetter_18} were able to obtain very good scaling beyond
$\mathord{\sim} 2000$ cores with the truncated multipole
expansion. This is conceivably due to the smaller amount of data that
needs to be exchanged in this method.  Data for comparison with other
methods for Cartesian grids are not readily available and difficult to
interpret because the resolution requirements for spherical and
Cartesian grids can differ significantly in computational stellar
astrophysics. We note, however, that, without careful optimization,
the 3D FFT on Cartesian grids of comparable size faces similar
scalability limits (see, e.g., the case with a grid of $128^3$ in
\citealt{eleftheriou_05}), so that the scalability of our algorithm
does not look excessively weak in comparison. Moreover, caution is in
order in comparing scaling measurements because we cannot account for
machine dependence.

For further optimization, one can
project onto functions of odd and even parity
in $\mu=\cos \theta$ before transforming from
$\tilde{s}_{i,\ell,m}$ to $\hat{s}_{i,j,m}$
and add the odd and even components
again after transforming from
$f_{i,\ell,m}$ to $g_{i,j,m}$. This
breaks up the multiplications with  $N_\theta\times N_\theta$ matrices
into two independent multiplications
with  $N_\theta/2 \times N_\theta/2$ matrices,
and roughly halves the amount of data
that needs to be sent to other MPI tasks.
This can help to speed up both
the serial
algorithm and the parallel algorithm
(for a large number of tasks,
as shown in the right column
of Table~\ref{tab:scaling})
by up to a factor of two. For
small parallel setups, the overhead from 
additional  point-to-point communication
can be counterproductive, however.

Especially when the solution is split into
odd and even components, the execution time is
sufficiently short for the algorithm to be useful for 3D simulations
that are dominated by other expensive components (e.g.\ microphysical
equation of state, nuclear burning, or neutrino transport).  Even in the \textsc{CoCoNuT-FMT} code, where the Poisson solver needs to be
called about 20 times for every update of the space-time metric,
simulations on $\mathord{\sim}1024$ cores remain feasible with the
linear solver consuming less than $20\%$ of the wall-clock time. More than half of
the wall-clock time of the non-linear metric
solver is still consumed outside
by other components, most notably the
recovery of the primitives.

\begin{figure}
    \centering
\plotone{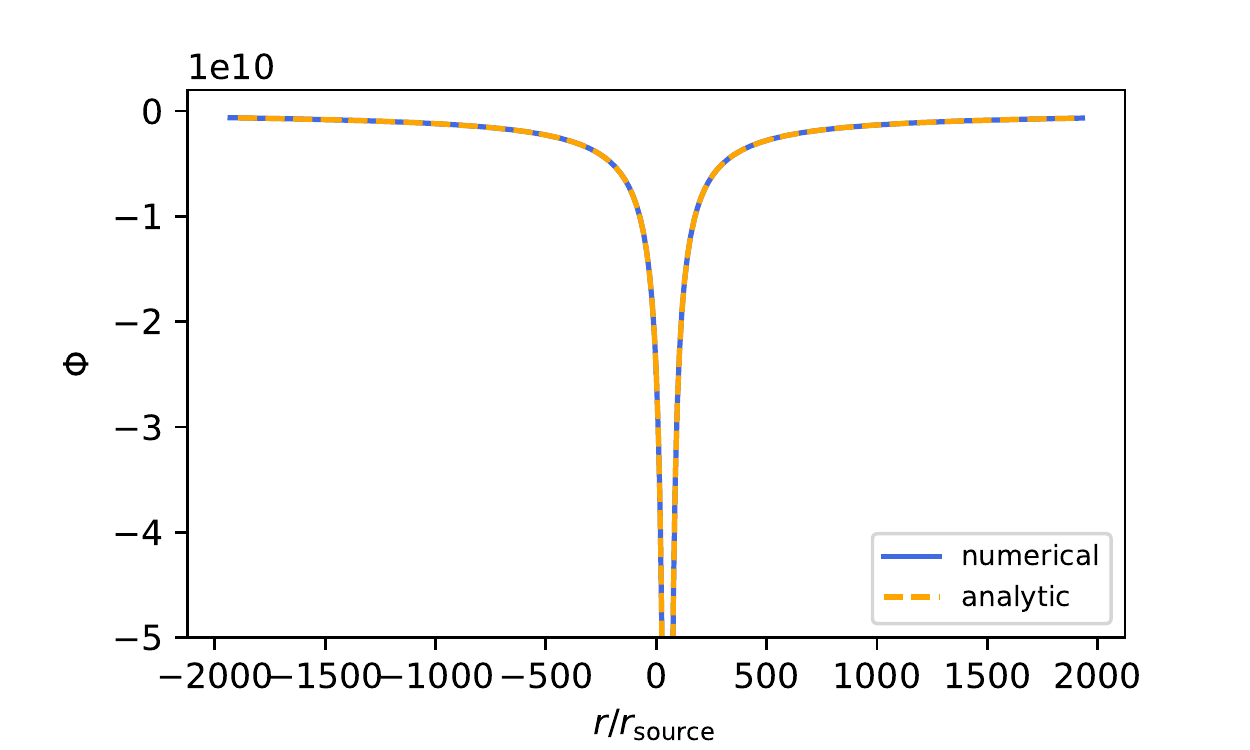}\\
\plotone{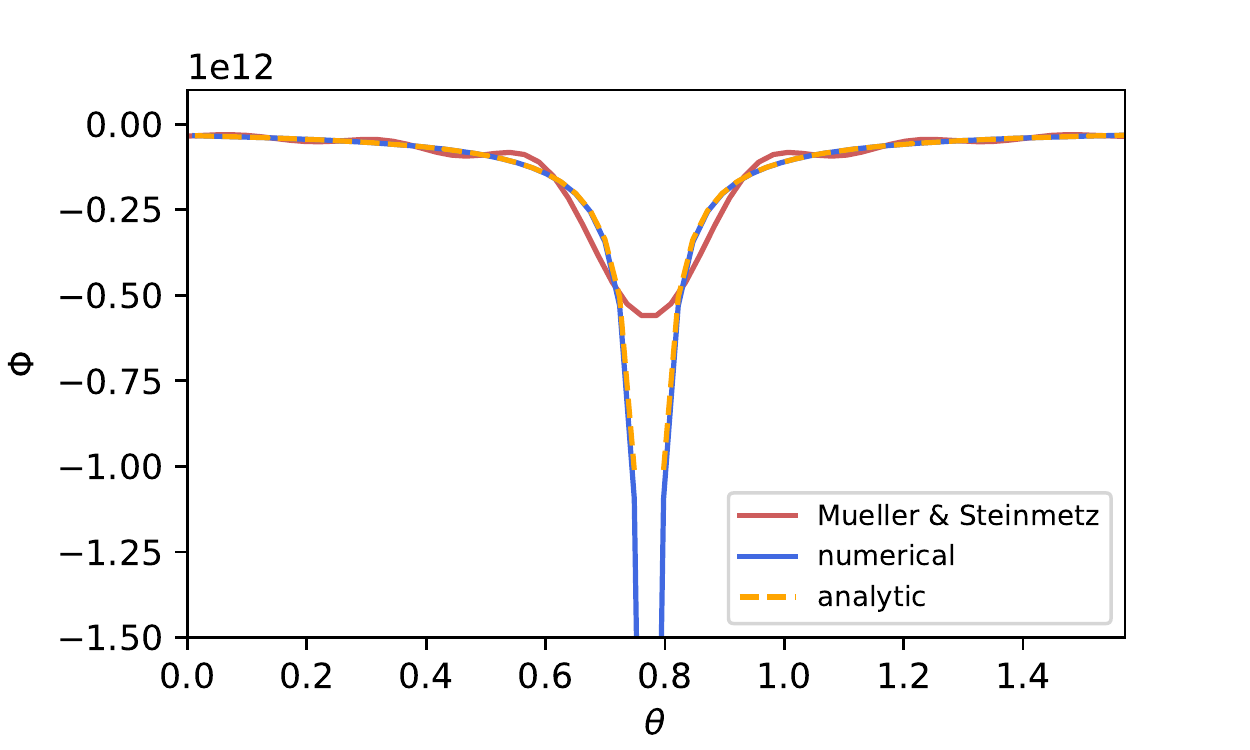}\\
\plotone{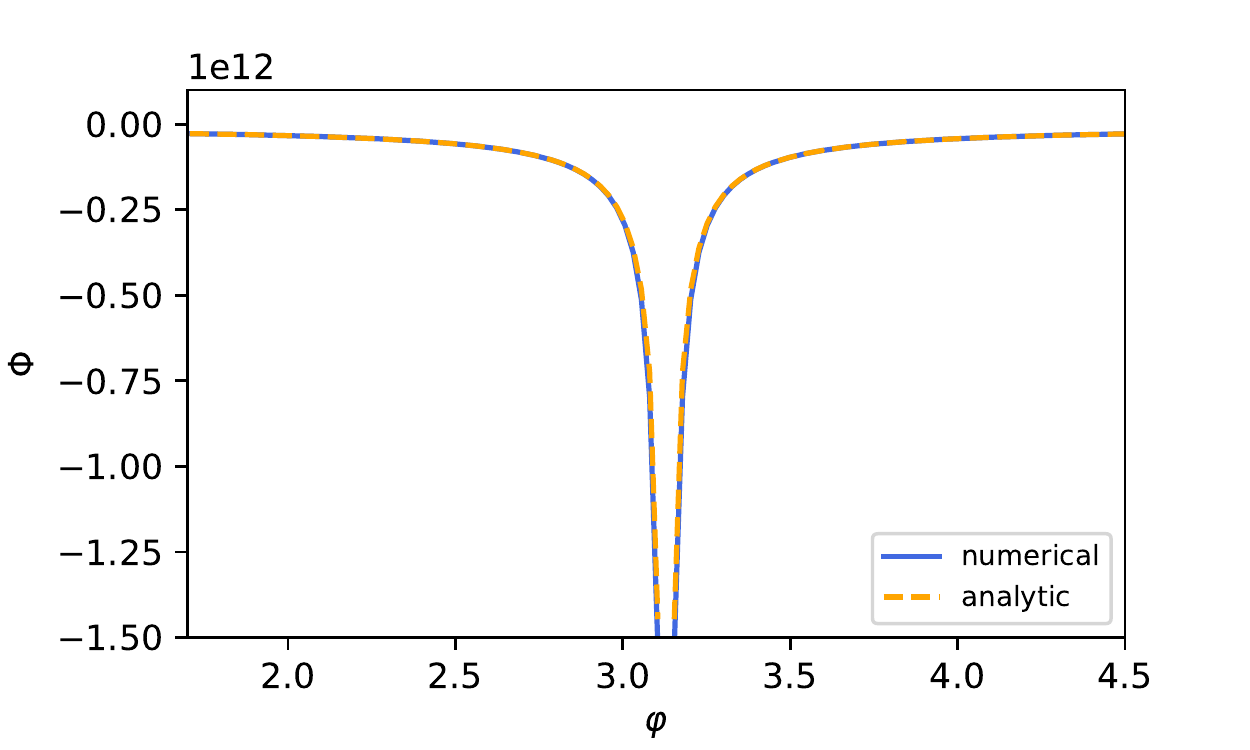}
\caption{Comparison of the analytic (dashed orange curves)
and numerical (blue solid curves) solutions for
the potential $\Phi$
of
a point source at $r=5.46 \times 10^7$,
$\theta=0.246 \pi$
and
$\varphi=0.996 \pi$ along the $r$-, $\theta$-
and $\varphi$-coordinate lines through
the source (top to bottom). In the middle panel,
we also show the result of the
algorithm of
\citet{mueller_95}
with a truncation point of
 $\ell_\mathrm{max}= 25$ for
 the multipole expansion (red curve).
 In the top panel, the radial coordinate
 is measured relative to the radial
 coordinate $r_\mathrm{source}$
 of the point mass.
    \label{fig:solutions}}
\end{figure}

\begin{figure*}
    \centering
\plottwo{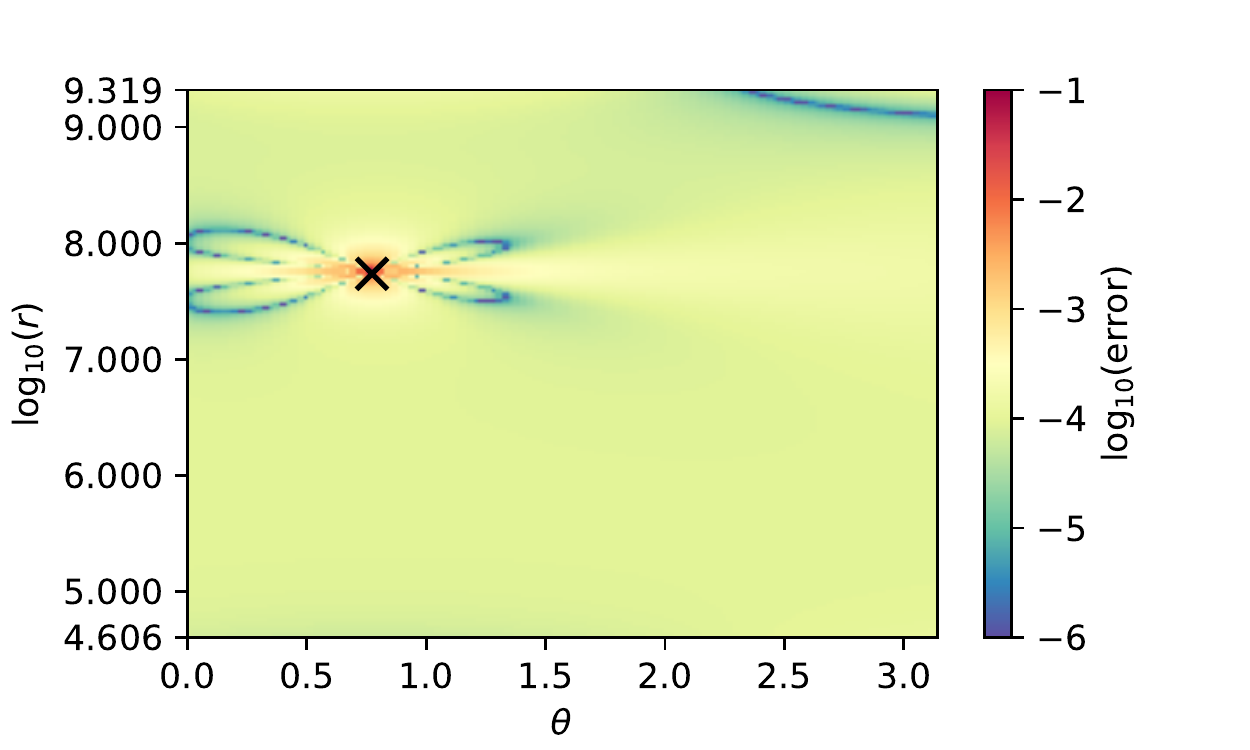}{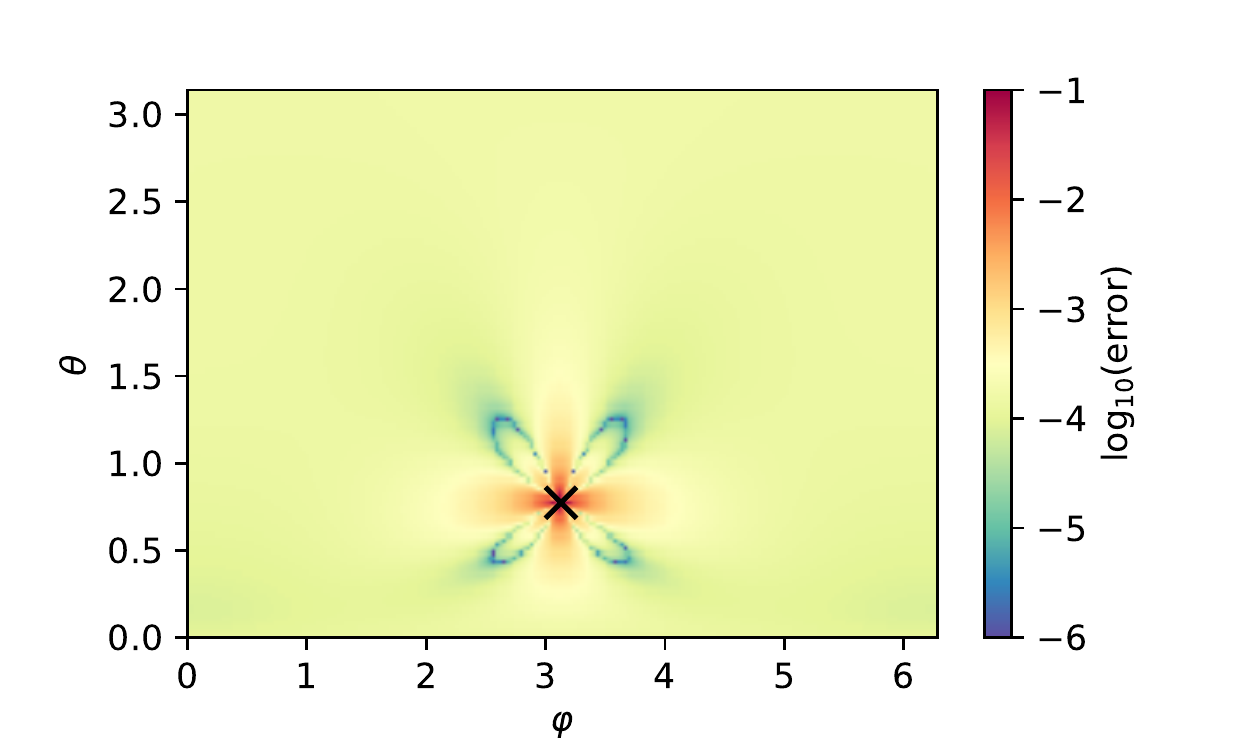}
\caption{Relative error of the numerical solution compared to
the analytic solution on coordinate slices
through the source 
with constant with constant
$\varphi$ (left) and constant $r$ (right). The position
of the point source is marked with a cross sign.
Note that
the numerical solution shows no visible artifacts
at the grid axis ($\theta=0$ and $\theta=\pi$).
    \label{fig:error}}
\end{figure*}

\begin{figure}
    \centering
    \includegraphics[width=\linewidth]{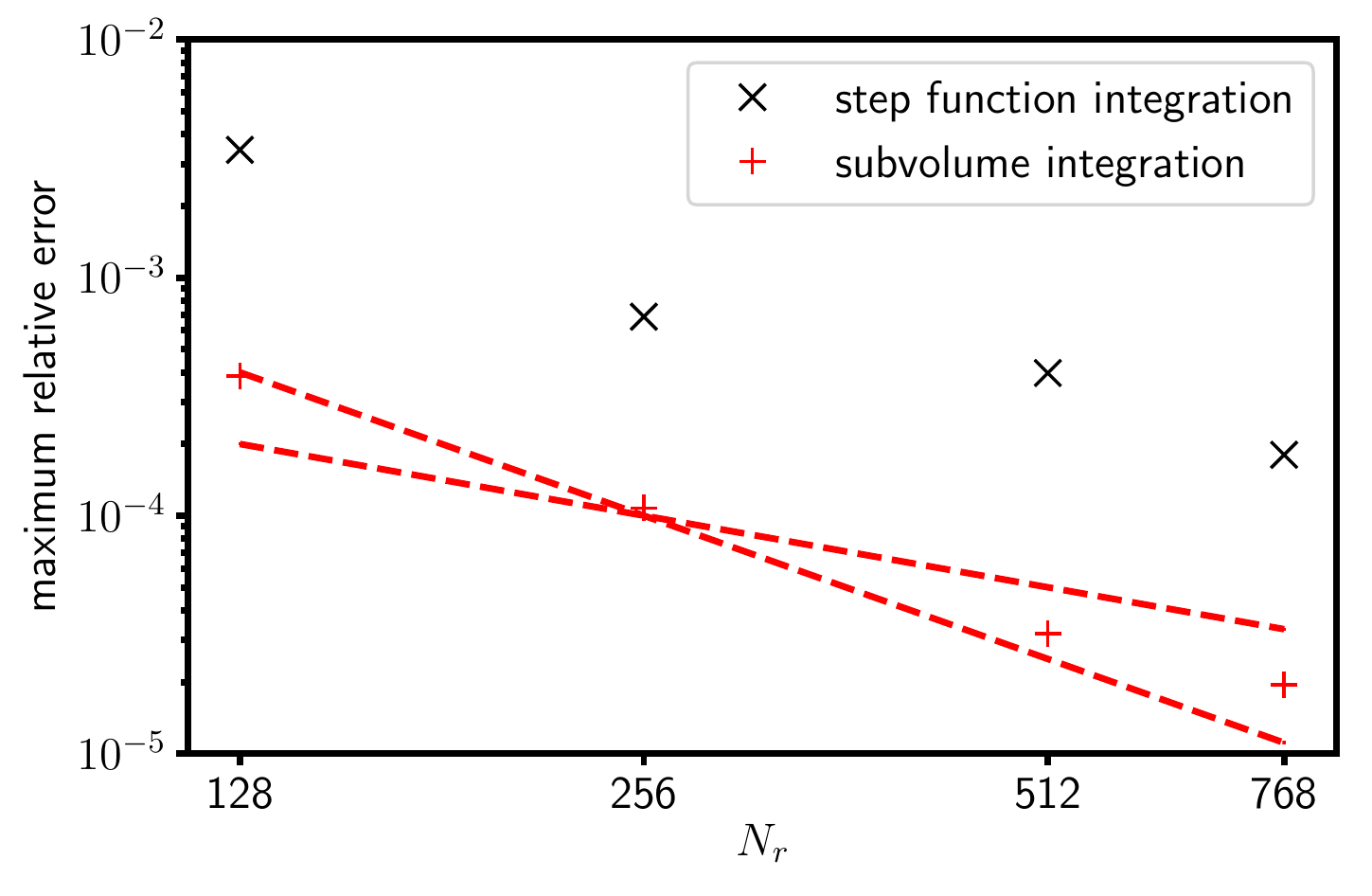}\\
    \includegraphics[width=\linewidth]{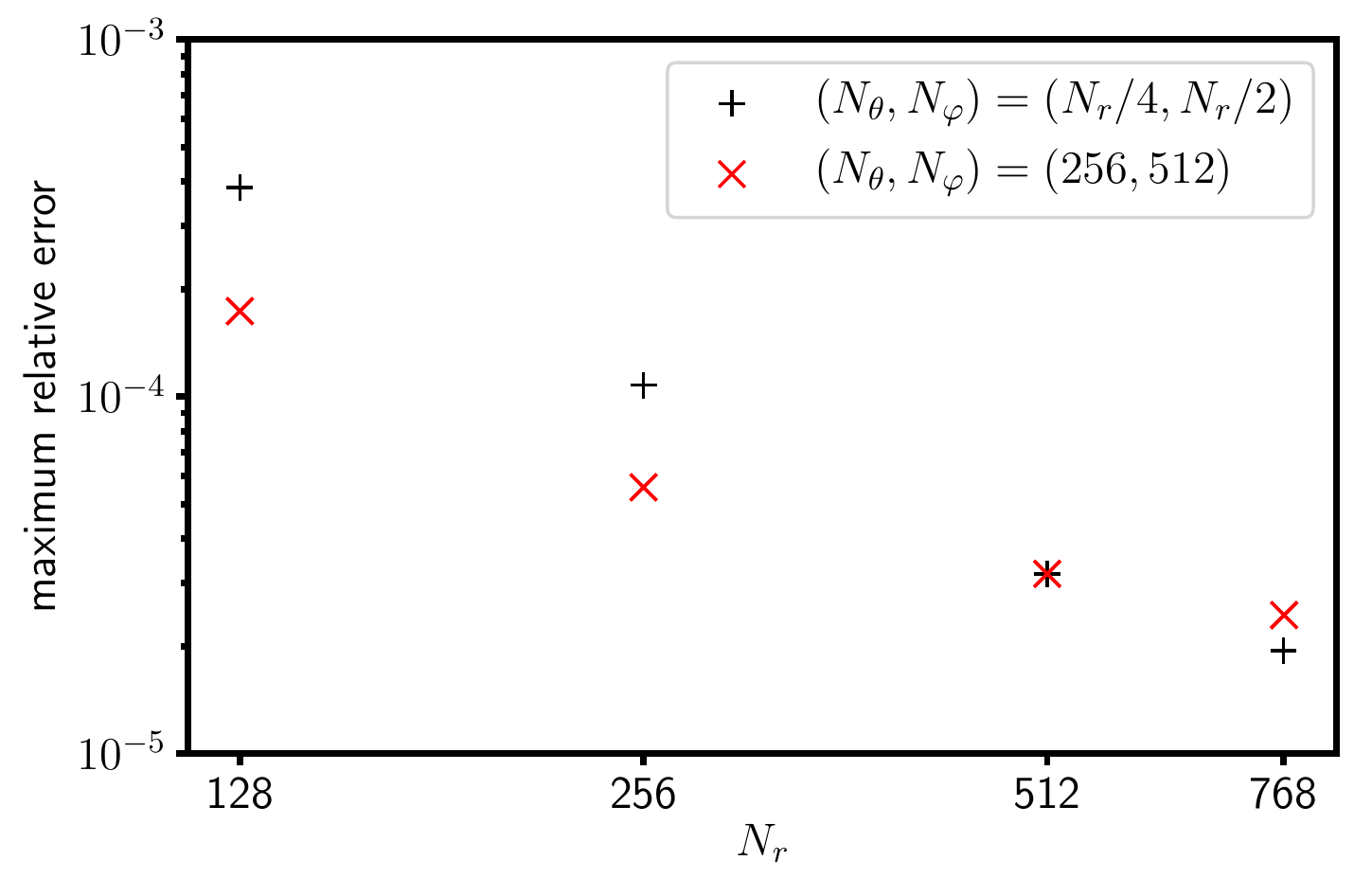}\\
    \includegraphics[width=\linewidth]{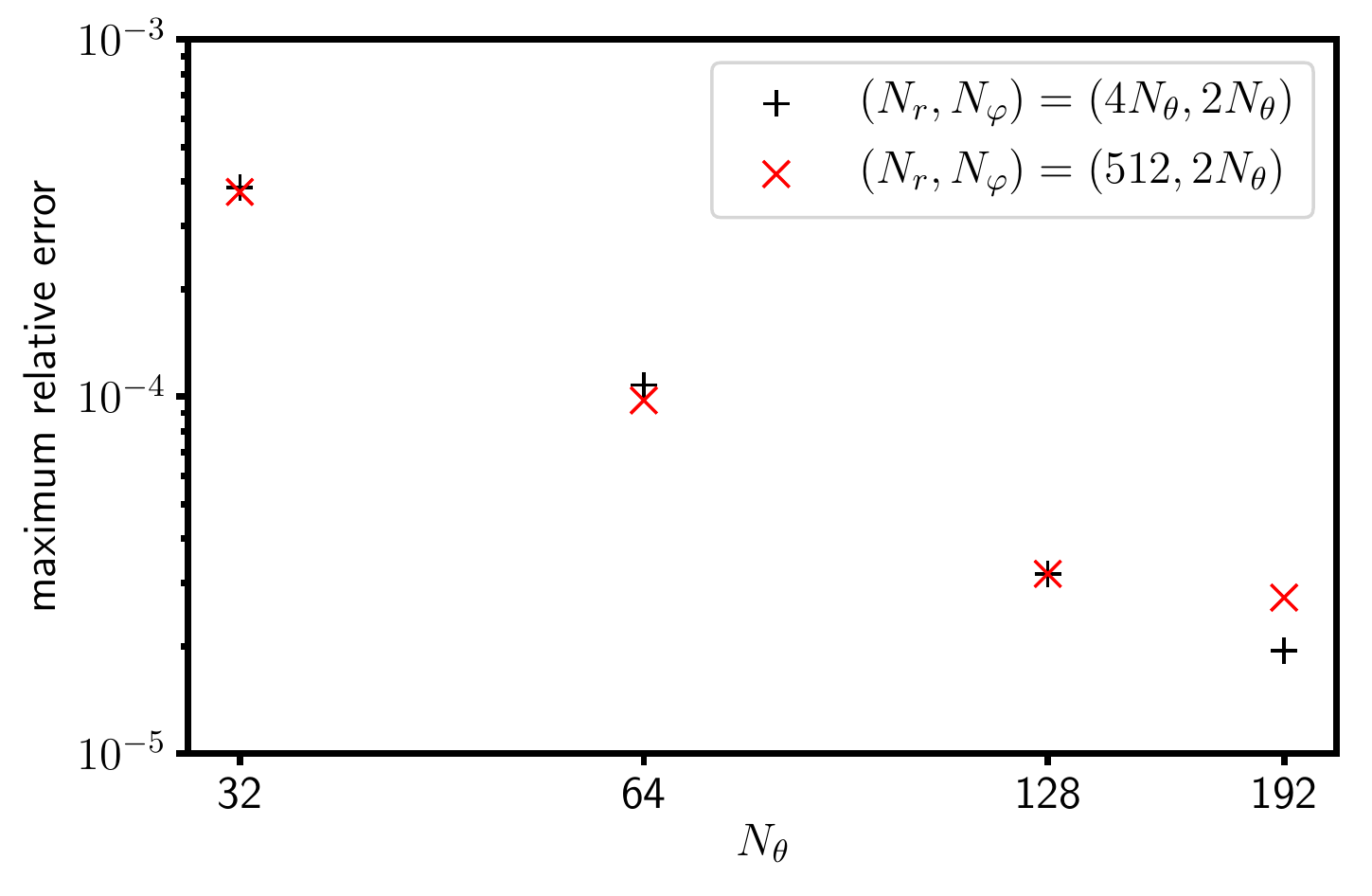}\\
    \caption{ Relative maximum error of the numerical solution for the
      case of a homogenous ellipsoid with $a=1$, $b=1.5$, and $c=2$. A
      uniform grid with an outer boundary of $r_\mathrm{out}=5$ is
      used in all cases.  The top panel shows the error as a function
      of grid size using a constant ratio
      $N_r:N_\theta:N_\varphi=4:1:2$.  The errors are larger if the
      mass per cell is evaluated using simple step function
      integration (black dots) and smaller by an order of magnitude if
      we use subvolume integration (red dots).  The dashed red lines
      indicate power laws of $N_r^{-1}$ and $N_r^{-2}$.  The middle
      panel compares the maximum error for varying $N_r$ using a
      constant ratio $N_r:N_\theta:N_\varphi=4:1:2$ (black) to
      solutions with varying $N_r$ and $N_\theta=128$ and
      $N_\varphi=512$ (red); subvolume integration is used in both
      cases.  The bottom panel compares the maximum error for varying
      $N_\theta=N_\varphi/2$ using a constant ratio
      $N_r:N_\theta:N_\varphi=4:1:2$ (black) to a series with
      $N_r=512$ (red).
    \label{fig:convergence}}
\end{figure}

\section{Verification}
\label{sec:verification}
It is customary to gauge approximate solvers
for the Poisson equation by comparing
to analytic solutions for configurations
with an extended density distribution,
such as MacLaurin spheroids
\citep{chandrasekhar_87} and various
axisymmetric disk models
\citep[e.g.][]{kuzmin_56,miyamoto_75,satoh_08}.

\subsection{Displaced Point Source}
In our case, we can also consider a much more
stringent test case, namely the field
of a point source displaced from the origin
of the grid. For code verification,
we choose a grid
with $550$ radial zones with
constant spacing in $\log r$
from $r=10^4$ 
to $r=2.1 \times 10^9$
(in non-dimensional units) and
$128\times 256$
uniformly spaced zones
zones in the $\theta$- and $\varphi$-direction.
A mass
of $m=1.352735 \times 10^{18}$
is placed in the zone with
indices $(i,j,k)=(366,32,128)$
or 
$r=5.46 \times 10^7$,
$\theta=0.246 \pi$
and
$\varphi=0.996 \pi$; this choice
corresponds to a density of $\rho=1$ in that zone.

Figure~\ref{fig:solutions} compares
the numerical solution to the analytic
solution for a point source along
three coordinate lines through the source,
and Figure~\ref{fig:error} shows
the relative error
on two surfaces with $\phi=\mathrm{const}.$ and
$r=\mathrm{const}.$ that intersect
the source location. Our solver
tracks the analytic solution almost
perfectly; even in the zones directly
adjacent to the point source, the maximum
relative error is only 10\%. The
Gibbs phenomenon that affects
the truncated multipole solver
of \citet{mueller_95}
(middle panel of Figure~\ref{fig:solutions})
is completely eliminated. Although the
Gibbs phenomenon
is absent or much less pronounced
in case of the truncated multipole expansion
for smoother, more extended sources, one must bear
in mind that the relative error in the potential
(which is typically used for the verification
of Poisson solvers; see
\citealp{mueller_95,couch_13b,almanstoetter_18})
can give a too favourable impression of
the solution accuracy. When the solution is used
to compute gravitational acceleration terms
in a hydrodynamics code, it is the \emph{derivatives}
of the potential
that matter, and these are much more severely
affected by the Gibbs phenomenon of the multipole
expansion than the potential itself.

\subsection{Convergence -- Potential of an Ellipsoid}
Since the analytic solution for the potential of
a point source is singular, this case is not well
suited for studying the convergence of the algorithm.
We therefore consider the potential
of an ellipsoid \citep{chandrasekhar_87} to address 
the convergence properties of the scheme; i.e.\
we use a source density of the form
\begin{equation}
\rho=
\left\{
\begin{array}{ll}
    1, &\frac{x^2}{a^2}+\frac{y^2}{b^2}+\frac{z^2}{c^2}\leq 1
    \\
    0, &\mathrm{else}
\end{array}
\right..
\end{equation}
The solution for $\Phi$ is given by \citep{chandrasekhar_87,almanstoetter_18}
\begin{equation}
    \Phi=
    \pi abc
    \left(A(\mathbf{r})X^2
    +B(\mathbf{r})y^2
    +C(\mathbf{r})z^2
    -D(\mathbf{r})\right),
\end{equation}
in terms of the integrals
\begin{eqnarray}
A(\mathbf{r})
&=&
\int\limits_u^\infty
\left[
(a^2+\lambda)^3(b^2+\lambda)(c^2+\lambda)
\right]^{-1/2}
\ud \lambda,
\\
B(\mathbf{r})
&=&
\int\limits_u^\infty
\left[
(a^2+\lambda)(b^2+\lambda)^3(c^2+\lambda)
\right]^{-1/2}
\ud \lambda,
\\
C(\mathbf{r})
&=&
\int\limits_u^\infty
\left[
(a^2+\lambda)(b^2+\lambda)(c^2+\lambda)^3
\right]^{-1/2}
\ud \lambda,
\\
D(\mathbf{r})
&=&
\int\limits_u^\infty
\left[
(a^2+\lambda)(b^2+\lambda)(c^2+\lambda)
\right]^{-1/2}
\ud \lambda.
\end{eqnarray}
Here $u$ is the real root
of
\begin{equation}
    \frac{x^2}{a^2+u}+\frac{y^2}{b^2+u}+\frac{z^2}{c^2+u}=1,
\end{equation}
if $\mathbf{r}$ lies outside the spheroid, and
$u=0$ otherwise. We choose an ellipsoid of
the same shape as \citet{almanstoetter_18} with
$a=1$, $b=1.5$, and $c=2$.

In Figure~\ref{fig:convergence}, we show the convergence of the
solution -- quantified by the maximum relative error -- on a uniform
grid in $r$ with an outer boundary of $r_\mathrm{out}=5$.  When
comparing the numerical and the analytic solution, we face two
separate issues: We introduce errors by discretizing $\Phi$ and
assuming constant source density within cells in a finite-volume
approach according to
Equations~(\ref{eq:fin_vol0}-\ref{eq:fin_vol3}). Moreover, errors in
the evaluation of the mass per cell and hence of the source density in
Equation~(\ref{eq:fin_vol0}) will also degrade the solution.  We deal
with those two types of discretization error by computing the source
in two ways. In the first approach, we use simple step-function
integration and set $\rho=1$ or $\rho=0$ depending on whether the cell
centre lies inside or outside the ellipsoid. In the second approach of
subvolume integration, we recursively divide the original cell volume
$\delta V$ into octants by halving the grid spacing
in $r$, $\theta$, and $\varphi$  before applying step-function integration.
The division into subvolumes is stopped if a refined cell either
lies completely inside or outside the ellipsoid, or if its
volumes is smaller than $10^{-6} \delta V$, where $\delta V$ is the
volume of the parent cell on the original grid. In both cases, we vary
the angular resolution from $N_r=128$ to $N_r=768$, keeping the ratio
$N_\varphi/N_r=2N_\theta/N_r=1/2$ constant.

The top panel of Figure~\ref{fig:convergence} shows that the
  evaluation of the source density completely dominates the error for
  this test problem. When the source density is evaluated accurately
  using subvolume integration, the error decreases slightly slower
  than $N_r^{-2}$. Since \citet{almanstoetter_18} do not specify
  whether they obtain the source density by step function integration
  or by a more accurate method, a direct comparison with their work
  using the truncated multipole expansion is difficult, but we note
  that we already obtain better accuracy for $N_r=512$ and
  $\delta_\theta=\delta_\varphi= 1.4^\circ$ than in their
  high-resolution case with $N_r=800$ and
  $\delta_\theta=\delta_\varphi= 1^\circ$.

The radial and angular resolution are not always of equal importance
for the solution error. In the middle and bottom panels of
Figure~\ref{fig:convergence}, we also show the maximum relative error
for cases with varying $N_r$ and constant $N_\theta=128$ and
$N_\varphi=256$ (middle panel) and with varying $N_\theta=N_\varphi/2$
and constant $N_r=512$. For this particular problem, decreasing
angular resolution with respect to our baseline case of
$(N_r,N_\theta,N_\varphi)=(512,128,256)$ appears to be much more
problematic than decreasing radial resolution.

We also investigate whether the solution accuracy is strongly
sensitive to local variations in the radial grid resolution. Still
using subvolume integration, we solve the Poisson equation on a grid
with a resolution of $(N_r,N_\theta,N_\varphi)=(512,128,256)$.  We
choose the radial grid to be identical to the uniform grid used before
up to $r=1.5$ and maintain constant $\delta r/r$ outside, so that the
outer boundary is at $r_\mathrm{out}=48.7$. This actually decreases
the maximum relative error from $3.2 \times 10^{-5}$ in the baseline
model to $2.7 \times 10^{-5}$. We also considered a jump in grid
resolution with a sudden increase of the (otherwise uniform) grid
spacing $\delta r$ by $50\%$ outside $r=1.5$, which increases the
maximum error slightly to $3.6 \times 10^{-5}$.  This suggests that
the algorithm deals well with variations in grid spacing that are not
too extreme.

\section{Conclusions}
\label{sec:summary}

We have presented an exact, non-iterative solver for the Poisson
equation on spherical polar grids. Compared to the truncated multipole
expansion \citep{mueller_95} used in many astrophysical simulation
codes based on spherical polar coordinates, our method has a number of
attractive features. Solving the discretized Poisson equation exactly
allows one to implement the gravitational momentum and energy source
terms in a fully conservative manner, and ensures
well-behaved convergence with increasing grid resolution. The method
also adroitly handles off-centred mass distributions without the need
to move the center of the spherical harmonics expansion
\citep{couch_13b}, and even multiple density concentrations are not an
obstacle.  This comes at little extra cost, since the operation count
of the algorithm is competitive with the standard multipole expansion
for $N_\ell =10\ldots 20$ for typical 3D grid setups. The parallel
performance is sufficient for the algorithm to be used in
hydrodynamical simulations at least on a few hundreds of cores. Further optimization of the parallel algorithm
may still be possible, e.g.\ by exploiting
symmetries in the FFT for real input data to
reduce the communication volume.
We make a \textsc{Fortran} implementation of the
serial algorithm and an easily adaptable template
of an MPI parallel version available under
{\color{red}\url{https://doi.org/10.5281/zenodo.1442635}}.

Although the method presented here is both
accurate and efficient, it comes with less flexibility
in the choice of the grid setup than the
standard multipole expansion.  The parallel
code currently requires the dimension
of the $\theta$- and $\varphi$- grid to be
a power of two. This, however, is not a fundamental
restriction and could be remedied by using
more general algorithms for the parallel FFT
and matrix-vector multiplication. 
A more serious limitation is that the algorithm
cannot readily be generalized to overset
spherical grids
\citep{kageyama_04,wongwathanarat_10a}
or spherical grids with non-orthogonal
patches like the cubed-sphere
grid \citep{wongwathanarat_16}.
One option would be to map to an auxiliary
global spherical
polar grid for the Poisson solver. 
In a distributed-memory paradigm, the
amount of data that needs to be communicated
between tasks would only be 
$\mathcal{O}(4 N_r N_\theta N_\varphi)$
for bilinear interpolation, which would
not increase MPI traffic tremendously. On the
downside, the mapped solution would no longer 
fulfill the discretized finite-volume form of
the Poisson equation exactly on the original grid,
and hence a major advantage of the algorithm would
be lost.

There are, however, alternative solutions for
some of the problems that prompt the use
of multi-patch grids or
non-orthogonal spherical grids in the first
place. The problem of stringent CFL time
step constraints near the grid axis can also
be solved or mitigated by filtering schemes
\citep{mueller_15b}
or non-uniform spacing in the $\theta$-direction,
which our new method can easily accommodate.
In the future, we will investigate whether further
refinements of these techniques can also reduce other 
shortcomings of spherical polar grids such as flow 
artifacts near the axis.

\section*{Acknowledgments}
This work was supported by the Australian Research Council through an
ARC Future Fellowships FT160100035 (BM). CC was supported by an
Australian Government Research Training Program (RTP) Scholarship.
This research was undertaken with the assistance of resources from the
National Computational Infrastructure (NCI), which is supported by the
Australian Government.  It was supported by resources provided by the
Pawsey Supercomputing Centre with funding from the Australian
Government and the Government of Western Australia and under Astronomy
Australia Ltd's merit allocation scheme on the OzSTAR national
facility at Swinburne University of Technology.

\bibliography{paper}

\appendix
\section{Non-Orthonormality of Standard Multipole Expansion}

\begin{figure}
    \centering
    \plottwo{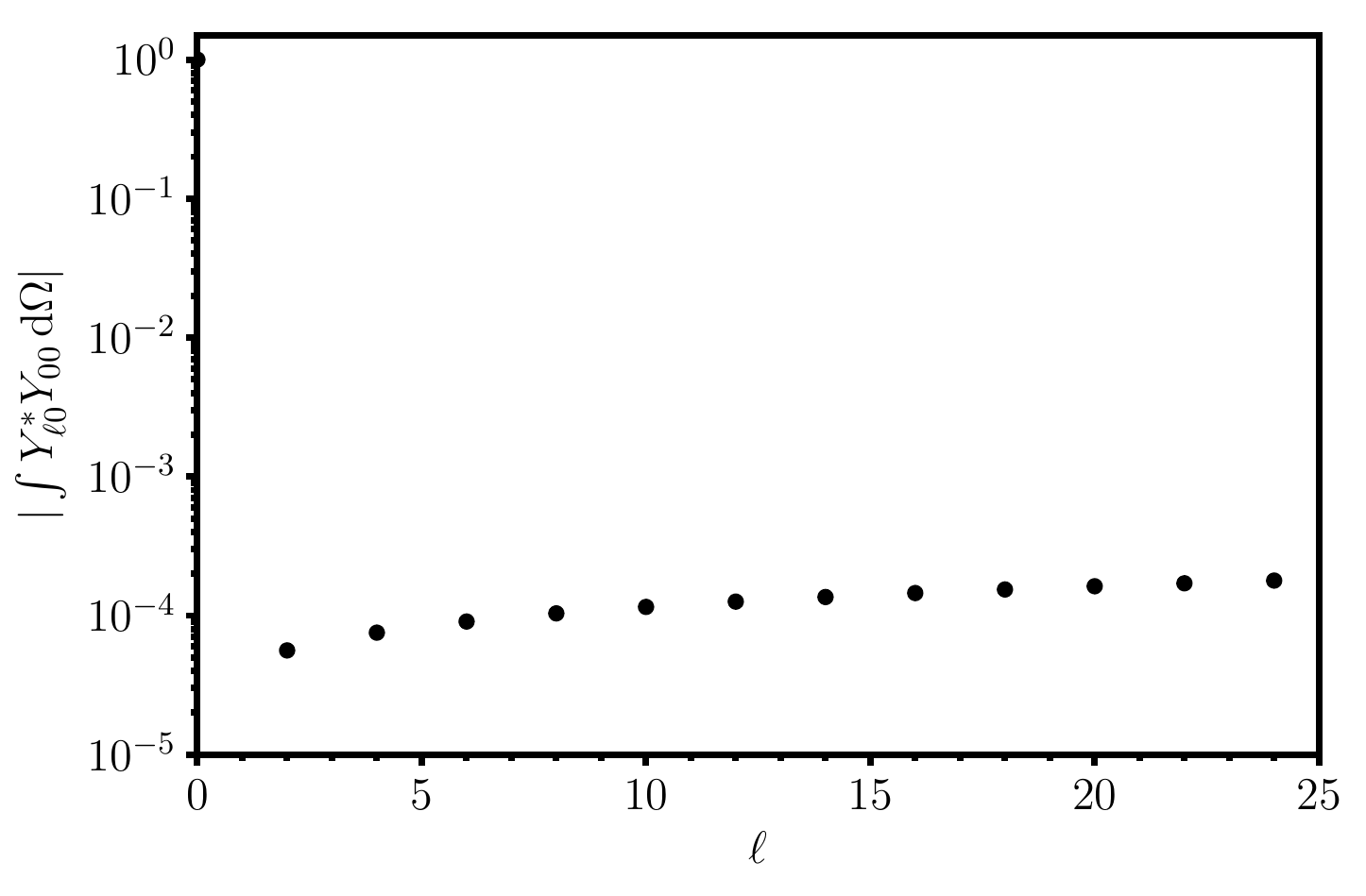}{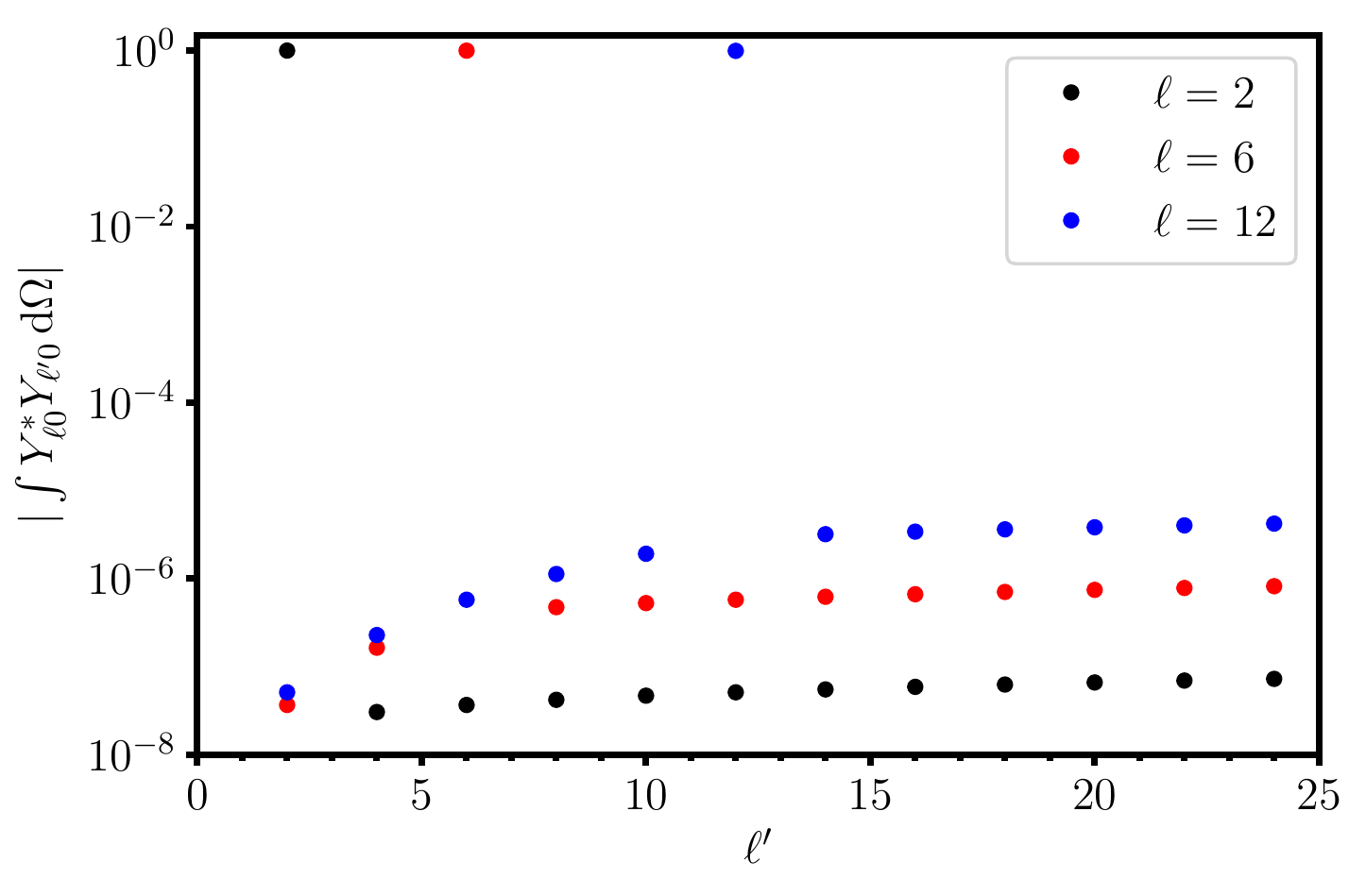}
    \caption{Left: Overlap between $Y_{00}$
    and spherical
    harmonics of even $\ell$  and $m=0$ for
    the cell-centred evaluation of spherical harmonics
    in the overlap integral in
    Equation~(\ref{eq:ovl}) in case
    of a uniform grid in $\theta$ with 128 zones.
    Right: Overlap between $Y_{2,0}$ (black),
    $Y_{6,0}$ (red), and $Y_{12,0}$ (blue)
    and other spherical
    harmonics of even $\ell'$  and $m=0$ 
    evaluated using the method of 
    \citet{mueller_95},
    again for a uniform grid in $\theta$ with 128 zones.
    \label{fig:ovl}}
\end{figure}

The use of standard spherical harmonics in the 
multipole expansion of the Green's function
(Equation~\ref{eq:green2}) can lead to solution
artifacts and divergence
problems if too many terms in the multipole expansion
are retained. While some of these problems can be
eliminated by using analytic integrals of
spherical harmonics \citep{mueller_95} or by
staggering the grids for the density and
the potential \citep{couch_13b}, other problems
are related to the failure of these methods
to respect the orthogonality relation
\begin{equation}
\int Y^*_{\ell m} Y_{\ell' m'} \,\ud \Omega
=\delta_{\ell \ell'}\delta_{m m'}.
\end{equation}

Let us consider the simplest method for decomposing
the source density into spherical 
harmonics in Equation~(\ref{eq:ode}).
If we already write the source density
as
$\rho(r,\theta,\varphi)=\sum_{\ell,m}c_{\ell m}(r) Y_{\ell m}(\theta,\varphi)$, and evaluate the overlap
integral using cell-centred values
for the spherical harmonics
as in \citet{couch_13b}, we obtain terms of the form
\begin{equation}
\label{eq:ovl}
    D_{\ell \ell,m m'}=\sum_{j,k}Y^*_{\ell m}(\theta_j,\varphi_k) Y_{\ell m}(\theta_j,\varphi_k) \Delta \Omega_{j,k},
\end{equation}
where $\Delta \Omega_{j,k}$ is the solid angle
occupied by a single cell.
In general, we will find
$ D_{\ell \ell,m m'} \neq \delta_{\ell \ell'}\delta_{m m'}$,
so that any multipole in the source gives rise to other spurious
multipoles in the solution. It is particularly problematic that even the monopole $Y_{00}$ overlaps with \emph{all}
other spherical harmonics with even $\ell$ and $m=0$
as illustrated in the left panel of Figure~\ref{fig:ovl}. 
Thus, the solution does not preserve spherical symmetry.
Moreover, the coefficients of the spurious multipoles
tend to be correlated so that they
manifest themselves
as small spikes near the axis that grow
and become narrower with larger $N_\ell$.
Whether or not the solution is evaluated
at cell centers or cell interfaces does not
change this behavior. The straightforward
cell-centred evaluation of spherical harmonics
in the overlap integral is therefore inadvisable
in spherical polar coordinates, although
it remains the only practical approach
in Cartesian coordinates.

The alternative approach of
\citet{mueller_95} merely assumes constant
source density within cells and then
evaluates the integrals over spherical
harmonics analytically. This is tantamount
to replacing
$Y_{\ell m}(\theta_j,\varphi_k)$
with its cell average
$\hat{Y}_{\ell m,jk}$,
\begin{equation}
Y_{\ell m}(\theta_j,\varphi_k)
\rightarrow
\hat{Y}_{\ell m,jk}
=\frac{1}{\Delta \Omega_{j,k}}
\int\limits_{\theta_{j-1/2}}^{\theta_{j+1/2}}
\int\limits_{\varphi_{k-1/2}}^{\varphi_{k+1/2}} 
Y_{\ell m} (\theta,\varphi) 
\sin \theta\,
\,\ud \varphi
\,\ud \theta
\end{equation}
in Equation~(\ref{eq:ovl}).
This ensures that overlap integrals with $Y_{00}$ reduce
to their correct analytic value so that $D_{\ell 0,m 0}=\delta_{\ell 0} \delta_{m 0}$ and the solution remains
spherically symmetric if the source density is. However,
the spurious overlap between higher multipoles is not
eliminated as shown in the right
panel of Figure~\ref{fig:ovl}. However, contrary to the naive
step-function integration, the spurious multipoles
do not pose serious problems for moderately large
values of the maximum multipole number $N_\ell$
in practice.

\end{document}